\documentclass{article}
\usepackage{graphics}
\usepackage{epsf}

\def\m#1{$#1$}
\def\tr{\;{\rm tr}\;}
\def\sgn{\;{\rm sgn}\;}

\newcommand{\beq}{\begin{equation}}
\newcommand{\eeq}{\end{equation}}

\newcommand{\beqs}{\begin{eqnarray}}
\newcommand{\eeqs}{\end{eqnarray}}

\newcommand{\half}{\frac{1}{2}}

\newtheorem{sutra}{}

\newtheorem{bhashya}{}[sutra]

\def\mbf#1{{\mbox{\boldmath $#1$}}}

\begin{document}
\bibliographystyle{h-physrev}
\input{epsf}

\title{A Theory Of Errors in Quantum Measurement}

\author{ S.G.Rajeev\thanks{rajeev@pas.rochester.edu} \\
University of Rochester. Dept of Physics and Astronomy. \\
Rochester. NY - 14627}
\maketitle

\begin{abstract}
It is common to model the random errors in a classical measurement by the
normal (Gaussian) distribution, because of  the central limit theorem.
In the quantum theory, the analogous hypothesis is that 
 the matrix elements of the error  in  an observable  are distributed normally. We 
obtain the  probability distribution this implies for
the outcome of a measurement, exactly for the case of traceless \m{2\times 2}
matrices and in the steepest descent approximation in general. Due to
the phenomenon of `level repulsion', the probability distributions
obtained are quite different from the Gaussian.

\end{abstract}

\pagebreak
In  classical physics,
 there is a well-established `standard model' of  random errors in measurement, the 
Gaussian or normal distribution \cite{eadie}: the  error is a sum of    a large number of
 more or less independent random contributions, so that the central
 limit theorem assures us that it  is a Gaussian. There is not as yet a
 similar theory of errors in quantum measurements. In the quantum
 theory, an observable  \m{\hat A} is  represented by a hermitian matrix. A simple
 model for the error would  again be that it is the sum of a large
 number of independent random additions to \m{\hat A}. If each matrix
 element is independent of the others (except for the condition of
 hermiticity),  the error would be described by a Gaussian random matrix
 \m{\hat B} added to the observable. A measurement of \m{\hat A} in the
 presence of this error would yield an eigenvalue of the sum \m{\hat
 R=\hat A+\hat B}
 instead of \m{\hat A}.

For such a random matrix \m{\hat R}, the analogue of the eigenvalue problem is to
ask for the probability distribution of the eigenvalues. Wigner solved such problems in the
context of nuclear physics \cite{wigner,porter}, followed by  fundamental
contributions by Dyson \cite{dyson}, Mehta \cite{mehta}, Itzykson and
Zuber \cite{iz}  and many others.  
By  now this theory has an
extensive literature in physics \cite{mehta} and mathematics \cite{voiculescu, katzsarnak}.
To make this paper more accessible, 
we will solve the problem of determining the probability distribution
of the eigenvalue  for a   simple example by elementary
means first. Then we turn to the more general case, using results from the
literature on  random matrix theory.

Let us recall in some more detail the theory of errors in classical physics.
An observable is a function
 \m{A:M\to R} from the phase space \m{M} taking  real
 values. The state of a classical system is given by a point \m{\xi\in M}
 in the phase space. Ideally, a measurement of the observable \m{A} on a
 system in this state will yield the value \m{A(\xi)}.The standard model
 of errors in classical measurement is 
 that there are a large number of  small, more or less  independent,  random
 corrections \m{B_1,\cdots B_M}  which add to this observable:
\beq
R=A+\sum_{k=1}^MB_r
\eeq 
Irrespective (upto mild assumptions) of the  distribution  of \m{B_1\cdots B_M}, the sum will,
 in the limit \m{M\to \infty} will tend to a Gaussian distribution (
 the central limit theorem)\cite{eadie}:
\beq
P(B)=e^{-cB^2}.
\eeq 

We
 can assume that the  mean of this distribution is zero since
 otherwise  it can be
 absorbed into the definition of \m{A}. ( In any case such systematic
 errors cannot be analyzed by  statistical methods.)
The variance  is usually assumed to be the same at all points in the 
phase-space so that \m{c} is a constant. Thus we model
 the outcome of the measurement  by a  Gaussian random variable whose mean is the `true' value
 \m{a=A(\xi) } of  the observable and the standard deviation is a measure of the size of
 the error:
\beq
p_A(x)\propto e^{-c(x-a)^2}.
\eeq
 There is by  now a well-developed sampling theory on how best
 to estimate this mean \m{a} and standard deviation \m{\sigma} from repeated
 measurements of \m{R} \cite{eadie,rao}.

If the state of the system is not known exactly, there is a
probability density function on the phase space \m{\rho:M\to R} which
determines the `instrinsic' probability distribution of \m{A} in the
absence of errors:
\beq
\tilde P_A(a)=\int \delta(a-A(\xi)) \rho(\xi)d\xi
\eeq
 If we add in the error, we get the convolution of
this with the Gaussian:
\beq
p_A(x)\propto \int \tilde P_A(x-B) P(B)  dB
\eeq

Thus there are two separate ways in which statistical considerations enter
classical error analysis: the classical observable can have an
intrinsic randomness because the state of the system is not completely
known; and there can be a random error added to the observable. In the quantum
theory, in addition there is another, more  fundamental source of 
randomness:  the uncertainty principle. Thus a proper theory of errors
has to take into account all three sources of randomness.

In the quantum theory, an observable \cite{vonneumann,bell}
 is  represented by a self-adjoint
 operator on the  Hilbert space of states. Let us 
 consider a hermitean  matrix
 \m{\hat A} of finite dimension \m{N}  with eigenvalues
\m{a_1,a_2\cdots a_N} and corresponding eigenstates \m{|u_1>\cdots |u_N>}.
Ideally, if a measurement of the observable represented by \m{A} is
made on a system known to be  in state \m{|\psi>}, the outcome will be one of the
 eigenvalues \m{a_k}, with probability \m{|<u_k|\psi>|^2}.
Another way to state this result (convenient in the following) is that
the probability density  function of this random variable  is 
\beq
\tilde P_{\hat A}(x)=<\psi|\delta(x-\hat A)|\psi>.
\eeq
More generally, if the state of the system is only  partially known,
 we have  a mixed state described by  a  density matrix
\m{\hat\rho} which is a positive hermitean matrix with \m{\tr \rho=1}.
We have then  the `intrinsic' probability
 distribution
\beq
\tilde P_{\hat A}(x)=\tr\hat\rho \delta(x-\hat A). 
\eeq
This formula takes into account two of the  sources of randomnes
 mentioned above: the fundamental uncertainty of quantum mechanics as
 well as our possible lack of knowledge of the state of the system.

However, this is still an ideal situation that is never realized in practice:
 there is always some error in the measurement which is a third source
 of randomness. It is useful to have
 theory of errors in quantum measurement analogous to the classical
 theory above,  particularly in
 view of current interest in quantum information theory.

As an elementary  example, suppose the observable is a traceless \m{2\times 2} matrix  
\m{\hat A=\pmatrix{a&0\cr 0&-a}}. We can think of it as the energy  of a spin half particle 
in a constant external magnetic field along the \m{z}-axis. There are
different sorts of errors that can affect this energy. The
simplest possibility is that  the magnetic
field fluctuates due, for example, to thermal effects in the currents
producing it. We can model
this as a sum of small independent additive corrections to the
magnetic field. 
By the central limit theorem, this additional magnetic field 
\m{\mbf{B}}  can be represented as a Gaussian random 
variable.
\beq
\hat R=\hat A+\mbf{\sigma}\cdot \mbf{B}.
\eeq
  The  mean of \m{\mbf{B}} can be assumed to be zero. Otherwise it can be absorbed into the
definition of the fixed magnetic field; after a rotation  we can
  reduce it to the above diagonal form anyway. It is reasonable, and
  simplest, to postulate that the error is  rotation invariant;i.e., all the components 
have the same variance:
\beq
 P(\mbf{B})\propto e^{-c\mbf{B}^2}.
\eeq
We can also think of this as the Boltzmann distribution for the
fluctuating magnetic field, as the energy of a magnetic field  is
proportional to \m{\mbf{B}^2}.
This means that the matrix elements of \m{\hat R} are also  Gaussian
random variables, with mean \m{\hat A}.

The eigenvalues of \m{\hat R=\hat A+ \mbf{ \sigma}\cdot \mbf{B}} will be  \m{\pm
  r}, where \m{r} is a real random variable. Calculating its    distribution  from that
of \m{\mbf{R}}, 
\beq
P_{\hat A}(\hat R)\propto e^{-\half c\tr [\hat R-\hat A]^2},
\eeq
is now the analogue of solving the eigenvalue problem. We will first
  consider the case where all the states are equally probable: the
  density matrix is a multiple of the identity. In effect we
  have to average over all fluctuations that change \m{\hat R} without
  affecting its eigenvalues.
Representing \m{\hat R=\mbf{\sigma}\cdot \mbf{R}} as a vector and transforming to spherical  
polar co-ordinates, we can perform the average over  random direction
  of the vector \m{\mbf{R}}:
\beqs
p_{\hat A}(r)&\propto& \int \tr \half\delta(r-\hat R)P_{\hat A}(\hat R)
  d\hat R\cr
&\propto& 
 r^2\int_{-1}^1
e^{-c(r^2+2ar\cos\theta)}\sin\theta d\theta\propto r\sinh[2acr]e^{-cr^2}.
\eeqs
Recalling that the observed eigenvalue can take also  negative values,
  we  normalize  this distribution to get 
\beq
p_{\hat A}(r)={1\over a}  \sqrt{c\over \pi}r\sinh[2acr]e^{-c(r^2+a^2)}.
\eeq
We can also write this in a way that explicity displays the peaks at
  \m{r=\pm a}:
\beq
p_{\hat A}(r)={1\over 2a}\sqrt{c\over \pi}r\left[e^{-c(r-a)^2}-e^{-c(r+a)^2}\right]
\eeq
In Fig. 1 we plot this probability distribution.

In summary,  the quantum mechanical error is modelled by a
Gaussian for the matrix elements which leads to a markedly
distribution from the Gaussian for the observed eigenvalue.
There is a second order zero for the distribution at the origin, a
consequence of the `level repulsion' of the eigenvalues of a random
matrix.  Note that the
peaks are displaced outwards from the `true' eigenvalues \m{\pm 1} due to
this level repulsion. (In the figures we have assumed  an unrealistically
large error to illustrate the phenomena better.)

The phenomenon of level repulsion is well-known in the theory of
random matrices. The set of all traceless hermitean 
\m{2\times 2} matrices with  a given spectrum \m{\pm r} is a
sphere of radius \m{r}. The volume of this sphere shrinks as \m{r\to
  0}, so that it is unlikely that the eigenvalues of a random matrix
are close together. We can also think of the logarithm of the volume
of the set of all matrices with a given spectrum as an entropy \cite{entropy}. The
probability distribution above can be thought of as  a  compromise between maximizing this
entropy and minimizing the energy of a fluctuation in \m{\mbf{B}}.

So far we dealt with   the case    when all the states of the system are
  equally likely;i.e,  when  the density  matrix is  proportional to the
  identity. The probability distribution in general will be given by
  averaging over the error as well as the states weighted by the
  density matrix:
\beq
 p_{\hat A}(x)=\int \tr\rho\delta(x-[\hat A+\hat B]) P(\hat B) dB
\eeq
In the special case of a two dimensional Hilbert space, we can expand
\beq
\hat \rho=\half +\mbf{\sigma}\cdot\mbf{\rho}, |\mbf{\rho}|\leq \half. 
\eeq
The inequality ensures that the density matrix is positive; the trace of \m{\hat \rho} is 
normalized to one.
We can again evaluate the integral by passing to spherical polar
  co-ordinates, using the identity
\beq
\delta(x-\mbf{\sigma}\cdot
  \mbf{R})=\half\left[\delta(x-r)+\delta(x+r)\right]+{\mbf{\sigma}\cdot\mbf{R}\over 2r}\left[\delta(x-r)-\delta(x+r)\right]
\eeq
to get 
\beqs
 p_{\hat A}(x)&\propto& \bigg\{x\left[e^{-c(x-a)^2}-e^{-c(x+a)^2}  \right]\cr
& & +{\eta\over a}\left[\left(x-{1\over 2ac}\right)e^{-c(x-a)^2}+\left(x+{1\over 2ac}\right)e^{-c(x+a)^2}
\right]\bigg\}
\eeqs
where \m{\eta=\tr \hat \rho \hat A} is the  average of \m{A} over states.
In Fig. 2 we plot this probability density in the extreme case of a
  system in an eigenstate of \m{A}. Note that there is a small  peak at the
  `wrong' eigenvalue, caused by the errors. 

We are ready now  to take up the more general case of  an \m{N\times N} hermitean matrix
 \m{\hat A} with an additive random error \m{\hat B}.
 We can assume that  the mean of \m{\hat B} is  the zero matrix: otherwise,
 we can redefine \m{\hat A} by absorbing this mean matrix into it.
For simplicity we assume that the matrix elements of \m{\hat B} are
 statistically independent of each other ( except for the constraint
 of hermiticity). In order for this condition to be true in all
 choices of basis, the joint probablity distribution should be a
 Gaussian \cite{mehta}:
\beq
P(\hat B)\propto e^{-c_1\tr \hat B^2-c_2(\tr \hat B)^2},
\eeq
for some positive  constants \m{c_1,c_2}.
Another justification of the normal distribution would come from
the central limit theorem: each matrix element of \m{B} is the
superposition of a  large number of small errors.

In the presence of the error, we are really measuring the observable
\m{\hat R=\hat A+\hat B}. The joint probability density function of its matrix elements
is
\beq
P_{\hat A}(\hat R)\propto e^{-c_1\tr [\hat R-\hat A]^2-c_2[\tr \hat
    R-\tr \hat A]^2}
\eeq
We can write \m{\hat R=\hat Ur\hat U^\dag} where \m{r} is a diagonal
matrix and \m{\hat U}
a unitary matrix. By averaging over \m{\hat U} we can get the joint
probability density of the eigenvalues of \m{\hat R}. In this process we
must remember that the Jacobian for transforming from the matrix
elements of \m{\hat R} to \m{r,\hat U} is \cite{mehta}
\beq
d\hat R\propto \Delta(r)^2d^Nrd\hat U, \quad \Delta(r)=\prod_{k<l}(r_k-r_l).
\eeq
Thus the joint distribution for the eigenvalues of \m{\hat R} is
\beq
p_{\hat A}(r_1,\cdots r_N)\propto \Delta(r)^2 e^{-c_1\sum_{k}\left[r_k^2+a_k^2\right]-
c_2\left[\sum_k r_k-\sum_ka_k\right]^2}\int d\hat U e^{2c_1\tr r\hat
  U^{\dag}\hat A\hat U}
\eeq
The last integral was evaluated by Harish-Chandra \cite{harish} ( and rediscovered
by Itzykson and Zuber \cite{iz} in a more physical context) 
\beq
\int_{U(N)}  d\hat U e^{2c_1\tr r \hat U\hat A\hat U^{\dag}}\propto 
{\det e^{2c_1r_k a_l}\over \Delta(r)\Delta(a)}.
\eeq
Thus
\beq
p_{\hat A}(r_1,\cdots r_n)\propto {\Delta(r)\over \Delta(a)}\det e^{2c_1r_ka_l}
e^{-c_1\sum_{k}\left[r_k^2+a_k^2\right]-c_2\left[\sum_k r_k-\sum_ka_k\right]^2}
\eeq
We can also write this result as
\beq
p_{\hat A}(r_1,\cdots r_N)\propto {\Delta(r)\over \Delta(a)}\sum_{\sigma\in
  S_N}
\sgn(\sigma)e^{-c_1\sum_k\left[r_k-a_{\sigma_k}\right]^2-
c_2\left[\sum_kr_k-\sum_ka_k\right]^2} 
\eeq
where the sum is over all permutations of the indices\m{\{1,2,\cdots
  N\}}.

For the  simplest  example   \m{A=\pmatrix{a&0\cr 0&-a}} we considered
earlier, we can write the above as 
\beq
p_{\hat A}(r_1,r_2)\propto e^{-[c_2+\half 
c_1]\left[r_1+r_2\right]^2}(r_1-r_2)\sinh[2ac_1(r_1-r_2)]e^{-\half c_1(r_1-r_2)^2}.
\eeq
This shows that the sum and difference of eigenvalues are independent
  random variables. The sum is a Gaussian random variable with mean zero. The
  difference has the distribution we derived earlier. Thus \m{c_2}
  describes a `classical' source of error which shifts both
  eigenvalues the same way  (and is a Gaussian) while \m{c_1} is a
  `quantum' error that affects their difference.

By integrating over all but one eigenvalue we can get 
the probability distribution for the outcome of a single measurement of
\m{A}. This leads to a complicated expression in general but in the
steepest descent approximation we can get a simpler form:
\beq
 p_{\hat A}(x)\propto \sum_{k=1}^N e^{-c(x-a_k)^2}\prod_{k\neq
  m}{|x-a_m|\over |a_k-a_m|},
\eeq
where \m{c=c_1+c_2}. This probability distribution is peaked near (but
not at) the eigenvalues \m{a_1,\cdots a_N} as one might expect. However even when
the density matrix is proportional to the identity, not all these
peaks are of the same height: due to the repulsion of eigenvalues of a
random matrix, the extreme values are more probable. Moreover, the peaks
of the probability are displaced towards the edges  due to this repulsion.
We plot an example in Fig. 3.

The spacing between eigenvalues are of order \m{\hbar}. In the limit
as \m{\hbar\to 0} and \m{N\to \infty} we  get back the classical
description. It is possible to develop a semi-classical theory along
the lines of Ref.\cite{neoclassical}. We hope to return to these
issues in a later publication.

It would be interesting to verify our predictions
experimentally. Perhaps Superconducting Quantum Interference Devices (
SQUIDs) or Spin Resonance will provide such tests. 

{\bf Acknowledgement}
I thank  A. Agarwal,G. Agrawal,  J. Eberly, G. Krishnaswami,
E. Langmann, A.C.  Melissinos  and C. Stroud for discussions.

\newpage

\centerline{\large\bf Figure Captions}
\medskip

\begin{enumerate}

\item Probablity density of the outcome of measuring an observable
  with \m{A} eigenvalues \m{\pm 1} in a mixed
  state with equal weight   for the  eigenvalues and error
  parameter \m{c=3}. The peaks are  shifted to  \m{\pm 1.15} from  \m{\pm 1}.

\item Probablity density of the same observable in an eigenstate with
  eigenvalue  \m{1.0} and error parameter \m{c=3}. The errors
  cause  a small peak at the `wrong' eigenvalue.

\item Probability density of an observable    with
  eigenvalues \m{-1,0.5,1.0 } with error parameter \m{c=2}. The peaks are
  shifted to \m{-1.25,0.36,1.41} respectively.

\end{enumerate}

\newpage
\vspace*{\fill}
\centerline{\scalebox{1.2}{\includegraphics{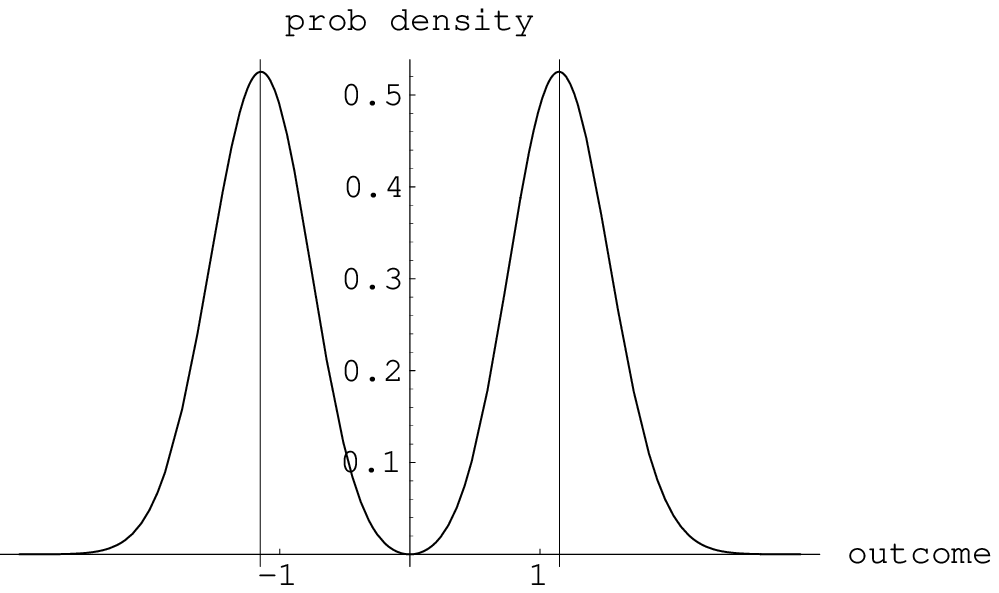}}}
\vfill
\centerline{Figure 1 of S. G. Rajeev}

\newpage
\vspace*{\fill}
\centerline{\scalebox{1.2}{\includegraphics{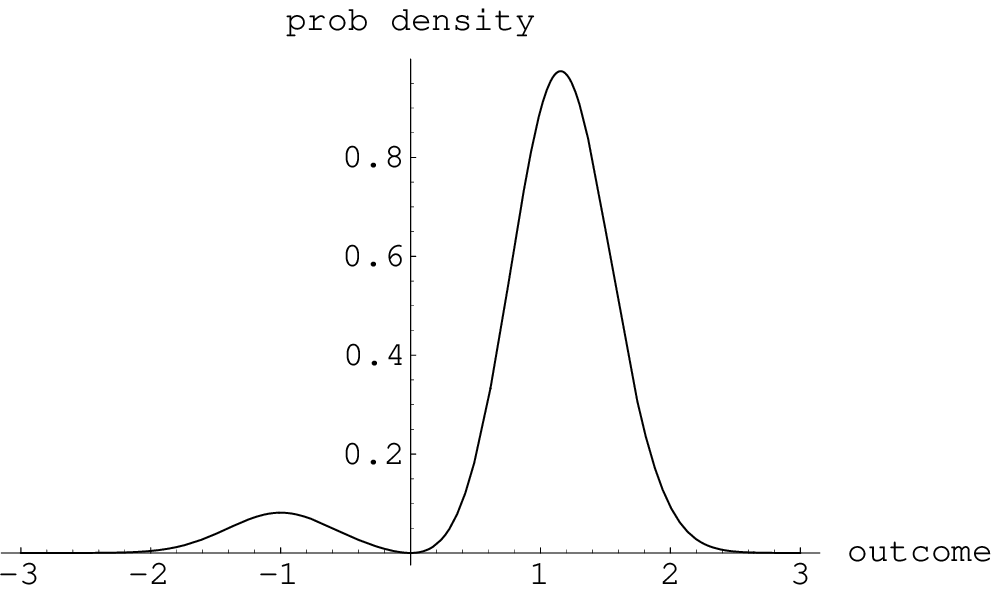}}}
\vfill
\centerline{Figure 2 of S. G. Rajeev}

\newpage
\vspace*{\fill}
\centerline{\scalebox{1.2}{\includegraphics{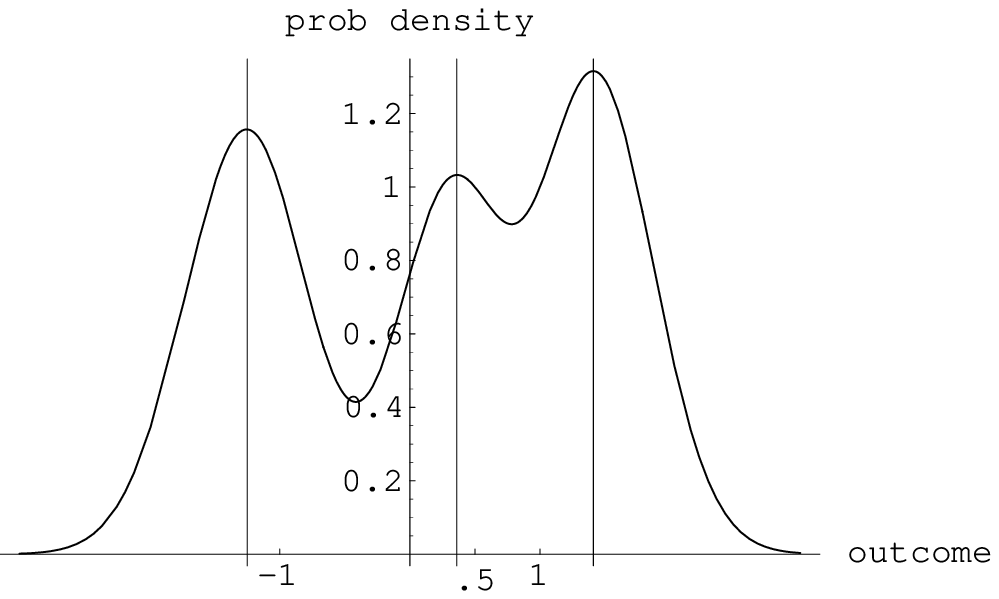}}}
\vfill
\centerline{Figure 2 of S. G. Rajeev}

\end{document}